\documentstyle[prb,aps]{revtex}

\def\al{\alpha}

\def\ga{\gamma}

\def\ve{\varepsilon}

\def\et{\eta}

\def\la{\lambda}

\def\fr#1#2{{{#1} \over {#2}}}

\def\half{{\textstyle{1\over 2}}}

\def\frac#1#2{{\textstyle{{#1}\over {#2}}}}

\def\lsim{\mathrel{\rlap{\lower4pt\hbox{\hskip1pt$\sim$}}
    \raise1pt\hbox{$<$}}}
\def\gsim{\mathrel{\rlap{\lower4pt\hbox{\hskip1pt$\sim$}}
    \raise1pt\hbox{$>$}}}
\def\sqr#1#2{{\vcenter{\vbox{\hrule height.#2pt
         \hbox{\vrule width.#2pt height#1pt \kern#1pt
         \vrule width.#2pt}
         \hrule height.#2pt}}}}

\def\etal{{\it et al.}}

\newcommand{\beq}{\begin{equation}}
\newcommand{\eeq}{\end{equation}}
\newcommand{\bea}{\begin{eqnarray}}
\newcommand{\eea}{\end{eqnarray}}
\newcommand{\rf}[1]{(\ref{#1})}

\newcommand{\bdm}{\begin{displaymath}}
\newcommand{\edm}{\end{displaymath}}

\begin{document}
\draft

\title{Dirac theory within the Standard-Model Extension}
\author{Ralf Lehnert}
\address{CENTRA, Departamento de F\'{\i}sica,
Universidade do Algarve,
8000-117 Faro, Portugal}
\maketitle
\begin{abstract}
The modified Dirac equation
in the Lorentz-violating Standard-Model Extension (SME)
is considered.
Within this framework,
the construction of a hermitian Hamiltonian
to all orders in the Lorentz-breaking parameters
is investigated,
discrete symmetries
and the first-order roots of the dispersion relation
are determined,
and various properties of the eigenspinors
are discussed.
\end{abstract}

\section{Introduction}
\label{intro}

Perhaps the most intriguing open question
in present-day fundamental physics
concerns a quantum theory of all fundamental interactions
including gravitation.
Experimental research in this field
is challenging
because quantum-gravitational effects
are expected to be suppressed
by the Planck scale $M_{\rm Pl}\simeq 10^{19}$ GeV.
However,
Lorentz violation is a promising candidate signature
of fundamental physics
that lies within the sensitivity range
of experiments with current or near-future technology.\cite{cpt01}

At presently attainable energy scales,
the effects of Lorentz violation
can be described within an effective-field-theory framework
called the Standard-Model Extension (SME).\cite{ck97,kl01,grav}
At the classical level,
the action of the SME
incorporates,
e.g.,
all leading-order contributions to the Lagrangian
that are formed by combining Standard-Model and gravitational fields
with Lorentz-breaking parameters
such that coordinate independence is maintained.
Nonzero parameters for Lorentz violation
can arise in a variety approaches to underlying physics
including strings,\cite{kps}
various non-string models of quantum gravity,\cite{foam}
noncommutative field theories,\cite{ncft}
varying couplings,\cite{vc}
random dynamics,\cite{rd}
multiverses,\cite{mv}
and brane-world scenarios.\cite{bws}
The flat-spacetime limit of the SME
has provided the basis for
numerous investigations of Lorentz violation
involving mesons,\cite{hadronexpt,hadronth,ak}
baryons,\cite{ccexpt,kl99,spaceexpt,cane}
electrons,\cite{eexpt,bkr,eexpt2,eexpt3}
photons,\cite{photonexpt,photonth,cavexpt,km}
muons,\cite{muons}
and neutrinos.\cite{ck97,neutrinos,nulong}

The extraction of the physical content of the SME
requires
an initial investigation
of the quadratic sectors of its Lagrangian
paralleling the conventional case.
More specifically,
the source-free equations of motion,
the associated Hamiltonians,
the dispersion relations,
and,
in the fermion case,
the eigenspinors
form a cornerstone
on which further theoretical studies
and comparisons with experiment rest.
For example,
the majority of the aforementioned analyses of Lorentz tests
involve the theory of free massive fermions of the SME.
Some basics
of this theory have been made plausible
or have been derived in certain limits as needed,
but a comprehensive treatment has been lacking.
The present work is intended to fill this gap.\cite{thesis}
More specifically,
we give a more rigorous and detailed study
of the general free Dirac equation and its solutions
in the context of the SME.
These results
provide important tools for further studies
of the physical implications of Lorentz violation.

The paper is organized as follows.
Section \ref{basics} sets up the notation,
reviews the basics of the modified Dirac equation,
and comments on its structure.
The construction
of a hermitian Hamiltonian
to arbitrary order in the Lorentz-violating parameters
is discussed in Sec.\ \ref{hamil}.
In Sec.\ \ref{disp},
we perform a systematic analysis
of discrete dispersion-relation symmetries.
Explicit leading-order approximations of the fermion eigenenergies
are obtained in Sec.\ \ref{1eigen}.
Sections \ref{spisym} and \ref{spipro}
contain an investigation of the eigenspinors
including a discussion of their symmetries
and a derivation of the generalized spinor projectors.
A brief summary
is presented in Sec.\ \ref{sum}.

\section{Basics}
\label{basics}

The general Lorentz-violating Lagrangian
for a single spin-$\frac{1}{2}$ fermion
\cite{ck97}
can be cast into a variety of forms.
One such form
reminiscent of the ordinary Dirac Lagrangian
and emphasizing the derivative structure
is \cite{kl99} 
\beq
{\cal L} = \frac{1}{2}{\it i}\overline{\psi}
{\Gamma}^{\nu}\hspace{-.15cm}
\stackrel{\;\leftrightarrow}
{\partial}_{\!\nu}\hspace{-.1cm}{\psi}
-\overline{\psi}M{\psi} \; ,
\label{lagr}
\eeq
where
\bea
{\Gamma}^{\nu} & \equiv & {\gamma}^{\nu}+c^{\mu \nu}
{\gamma}_{\mu}+d^{\mu \nu}{\gamma}_{5}
{\gamma}_{\mu}+e^{\nu}+if^{\nu}{\gamma}_{5}
+\frac{1}{2}g^{\lambda \mu \nu}
{\sigma}_{\lambda \mu}
\; ,\nonumber\\
\quad M & \equiv & m+a_{\mu}{\gamma}^{\mu}+b_{\mu}{\gamma}_{5}
{\gamma}^{\mu}+\frac{1}{2}H^{\mu \nu}
{\sigma}_{\mu \nu}
\label{Gam}
\; .
\eea
The gamma matrices $\{I, \gamma_5,\gamma^{\mu},
\gamma_5\gamma^{\mu}, \sigma^{\mu \nu}\}$
have conventional properties,
and the signature of the Minkowski metric
$\eta_{\mu\nu}$ is $-2$.
The Lorentz-violating parameters
$a_{\mu}$, $b_{\mu}$, $c_{\mu \nu}$,
$d_{\mu \nu}$, $e_{\mu}$, $f_{\mu}$,
$g_{\mu \nu \lambda}$, and $H_{\mu \nu}$
are taken as real
with
$c_{\mu \nu}$ and $d_{\mu \nu}$ traceless,
$g_{\mu \nu \lambda}$ antisymmetric
in its first two indices,
and $H_{\mu \nu}$ antisymmetric.
Note 
that $a_{\mu}$, $b_{\mu}$, $e_{\mu}$, $f_{\mu}$,
and $g_{\mu \nu \lambda}$
break CPT symmetry as well.
For definiteness,
we assume $m>0$.
However, 
many of our results
also hold in the massless case.
On phenomenological grounds, 
all Lorentz-violating coefficients
must be minute in a certain class
of inertial frames
called {\it concordant frames},
in which the Earth moves non-relativistically.
Then, 
paralleling the usual Dirac case, 
a hermitian Hamiltonian
with two positive and two negative eigenvalues 
exists.\cite{kl01}

For future reference,
we provide an equivalent form
of Lagrangian (\ref{lagr})
emphasizing its gamma-matrix structure:
\cite{kl01}
\beq
{\cal L} = \overline{\psi}
(S+iP\gamma_5+V^{\mu}\gamma_{\mu}
+A^{\mu}\gamma_5\gamma_{\mu}
+T^{\mu\nu}\sigma_{\mu\nu})
{\psi}
\label{lagr2}
\; .
\eeq
Here, we have defined
\begin{eqnarray}
& S[i\partial]\equiv e^{\mu}i\partial_{\mu}-m\; ,\quad\;
P[i\partial]\equiv f^{\mu}i\partial_{\mu}
\; ,\quad\;
V^{\mu}[i\partial]\equiv i\partial^{\mu}
+c^{\mu\nu}i\partial_{\nu}-a^{\mu}\; ,\quad\;
A^{\mu}[i\partial]\equiv d^{\mu\nu}
i\partial_{\nu}-b^{\mu}\; , & \nonumber\\
& T^{\mu\nu}[i\partial]
\equiv \frac{1}{2}(g^{\mu\nu\rho}
i\partial_{\rho}-H^{\mu\nu})\; , &
\label{lagr2ex}
\end{eqnarray}
where
the dependence of the introduced quantities
on the quantum-mechanical 4-momentum operator
has been displayed for clarity.
The Lagrangians (\ref{lagr}) and (\ref{lagr2})
differ only by a total divergence, 
so that they are physically equivalent. 
Note that the replacement
$i\partial\rightarrow\frac{1}{2}i\hspace{-.15cm}
\stackrel{\;\leftrightarrow}{\partial}$
in Def.\ (\ref{lagr2ex})
renders both Lagrangians identical.

The associated modified Dirac equation is given by 
\beq
(i\Gamma^{\mu}\partial_{\mu}-M)\psi(x)=0
\; .
\label{diraceq}
\eeq
A Klein-Gordon-type equation
can be obtained employing the following squaring procedure.
Consider the  modified Dirac operator
$(i\Gamma^{\mu}\partial_{\mu}-M)$ 
and change the signs of the parameters
$b^{\mu}$, $d^{\mu\nu}$, $g^{\mu\nu\rho}$,
and $H^{\mu\nu}$. 
Application of the resulting operator
to Eq.\ (\ref{diraceq})
yields the desired second-order equation
\beq
(\breve{S}+\breve{P}
\gamma^5+\breve{V}^{\mu}
\gamma_{\mu})\psi(x)=0
\; ,
\label{kgeq}
\eeq
where we have defined
\beq
\breve{S}\equiv S^2
-P^2+V^2
+A^2-2T^2 \; ,\quad
\breve{P}\equiv2(iPS-V\cdot A
-iT\tilde{T})
\;  ,\quad
\breve{V}^{\mu}
\equiv2(SV^{\mu}+iPA^{\mu}
-2iT^{\mu\nu}V_{\nu}+2\tilde{T}
^{\mu\nu}A_{\nu})
\; .
\label{abbrcof}
\eeq
In the above expressions,
the dependence of the various quantities 
on $i\partial$ is understood.
The tensor $\tilde{T}^{\mu\nu}
=\frac{1}{2}
\epsilon^{\mu\nu\alpha\beta}
T_{\alpha\beta}$
denotes the dual and
$\epsilon^{\mu\nu\alpha\beta}$
is the totally antisymmetric symbol
with $\epsilon^{0123}=+1$, as usual.
The unconventional
off-diagonal pieces in Eq.\ \rf{kgeq} 
can be eliminated 
with a second squaring procedure 
involving the application of 
$(\breve{S}-\breve{P}\gamma^5
-\breve{V}^{\mu}\gamma_{\mu})$
to the Klein-Gordon-type equation \rf{kgeq}.
As expected, 
the resulting fourth-order operator 
can be expressed as the determinant 
of the modified Dirac operator, 
paralleling the conventional case:
\beq
{\rm det}
(i\Gamma^{\mu}\partial_{\mu}-M)\psi(x)=0
\; .
\label{fourth}
\eeq
Thus, 
each individual component of a spinor 
solving the modified Dirac equation \rf{diraceq}
satisfies Eq.\ \rf{fourth}. 

A plane-wave ansatz
$\psi(x)=\exp(-i\lambda^{\mu}x_{\mu})
W(\vec{\lambda})$ 
for solutions to the modified Dirac equation \rf{diraceq}
yields
\beq
(\Gamma^{\mu}\lambda_{\mu}-M)
W(\vec{\lambda})=0
\label{spinordet}
\eeq
determining the four-component spinor
$W(\vec{\lambda})$,
where the 4-momentum $\lambda^{\mu}$
must solve the dispersion relation
\beq
{\rm det}
(\Gamma^{\mu}\lambda_{\mu}-M)=0
\; .
\label{disprel}
\eeq
With our earlier assumption,
dispersion relation (\ref{disprel})
has two positive-valued roots $\lambda^0_{+ (1,2)}(\vec{\lambda})$
and two negative-valued ones
$\lambda^0_{- (1,2)}(\vec{\lambda})$.
The corresponding
4-momenta and eigenspinors 
are $\lambda^{\mu}_{\pm(\alpha)}$
and $W^{(\alpha)}_{\pm}(\vec{\lambda})$, 
respectively.
Throughout this work,
indices in parentheses can take the values 1 and 2.
After the usual reinterpretation
of the negative-energy solutions,
the 4-momenta are denoted by
\begin{eqnarray}
p^{(\alpha)}_u\equiv\lambda_{+(\alpha)}
\; ,\quad\;
p^{(\alpha)}_v\equiv-\lambda_{-(\alpha)}
\; ,
\label{reinterpr1}
\end{eqnarray}
where we have omitted
the Minkowski indices
for brevity.
The notation for the reinterpreted eigenspinors is
\begin{eqnarray}
U^{(\alpha)}(\vec{p})\equiv
W^{(\alpha)}_{+}(\vec{\lambda})
\; ,\quad\;
V^{(\alpha)}(\vec{p})\equiv
W^{(\alpha)}_{-}(-\vec{\lambda})\; .
\label{reinterpr2}
\end{eqnarray}
The spinors
and the dispersion relation 
are discussed in more detail
in subsequent sections.

For a 4-momentum $\lambda_{\mu}$
that fails to satisfy
dispersion relation
(\ref{disprel}),
the cofactor matrix
of the modified Dirac operator $(\Gamma^{\mu}\lambda_{\mu}-M)$
in $\lambda$-momentum space
is given by
\beq
{\rm cof}(\Gamma^{\mu}\lambda_{\mu}-M)
=\det (\Gamma^{\mu}\lambda_{\mu}-M)
(\Gamma^{\mu}\lambda_{\mu}
-M)^{-1}
\; .
\label{cofdef}
\eeq
This matrix
appears in many applications
of our model (\ref{lagr}),
such as the anticommutator function.\cite{kl01}
A more explicit expression is therefore desirable. 
As a corollary
of the above discussion
of the equations of motion,
we obtain 
\beq
{\rm cof}(\Gamma^{\mu}\lambda_{\mu}-M)
=(\breve{S}-\breve{P}\gamma^5
-\breve{V}^{\mu}\gamma_{\mu})
(S+iP\gamma_5+V^{\mu}\gamma_{\mu}
-A^{\mu}\gamma_5\gamma_{\mu}
-T^{\mu\nu}\sigma_{\mu\nu})
\; .
\label{cofdef2}
\eeq
As we are now working in momentum space,
the according change from $i\partial$
to $\lambda$ in Defs.\ \rf{lagr2ex} and \rf{abbrcof}
is implied.
Replacing $\lambda\rightarrow i\partial$ everywhere
in (\ref{cofdef2}),
yields the associated relation
in position space, as usual.

\section{Hamiltonian}
\label{hamil}

In concordant frames, 
$\Gamma^0$ is invertible,\cite{kl01}
so that the modified Dirac equation
(\ref{diraceq}) can be cast
into Schr\"odinger form:
\beq
i\partial_0\psi(x)
=(\Gamma^0)^{-1}(i\vec{\Gamma}
\cdot\vec{\nabla}+M)\psi(x)
\; .
\label{schrodingereq}
\eeq
Although the operator
$\tilde H[i\vec{\nabla}] \equiv
(\Gamma^0)^{-1}(i\vec{\Gamma}
\cdot\vec{\nabla}+M)$
appearing above
is reminiscent of a Hamiltonian,
it fails to be hermitian in general.
This results in such undesirable features
like a non-unitary time evolution.
This issue can be resolved
by a spinor redefinition
$A\chi\equiv\psi$
chosen to eliminate
the unconventional time-derivative couplings.\cite{bkr}
Here, $A$ is a non-singular
spacetime-independent
4$\times$4 matrix,
which exists in concordant frames.\cite{kl01}
This field redefinition
leaves unaffected the physics
because it is a canonical transformation.
It amounts
to a change of basis in spinor space.

The existence of $A$ is equivalent\cite{kl01} 
to the positive definiteness of $\gamma^0\Gamma^0$.
This permits the definition\cite{lt}
of a unique, positive-definite, invertible matrix $(\gamma^0\Gamma^0)^{1/2}$.
The matrix $A$ can then be expressed as
\beq
A=(\gamma^0\Gamma^0)^{-1/2}
\; .
\label{matrixa}
\eeq
Note that the hermiticity of $\gamma^0\Gamma^0$
yields $A=A^{\dagger}$.
It can now be verified
that the Hamiltonian $H$
given by
\beq
H[i\vec{\nabla}]
=(\gamma^0\Gamma^0)^{-1/2}\gamma^0
(i\vec{\Gamma}\cdot\vec{\nabla}+M)
(\gamma^0\Gamma^0)^{-1/2}
\; ,
\label{hamiltonian}
\eeq
which governs the time evolution
of the redefined field $\chi$,
is hermitian,
as desired.
We also point out
that $H$ and $\tilde{H}$
are related by the similarity
transformation
\beq
H=(\gamma^0\Gamma^0)^{1/2}\tilde{H}
(\gamma^0\Gamma^0)^{-1/2}
\; .
\label{simtrans}
\eeq

The determination of the explicit form of the matrix $A$
is challenging
in the general.
In practice,
however,
it usually suffices
to determine $A$
up to a given order
in the Lorentz-violating coefficients.
Notice that $\gamma^0\Gamma^0$
can be split into the 4$\times$4 identity $I$
plus a Lorentz-breaking correction:
$\gamma^0\Gamma^0=I+\gamma^0(\Gamma^0
-\gamma^0)$.
This suggests the following expansion:
\beq
A=I+\sum_{n=1}^{\infty}
\frac{\displaystyle (2n-1)!!}
{\displaystyle (2n)!!}
\left(I-\gamma^0\Gamma^0\right)^n
\; .
\label{aexpansion}
\eeq
In a basis in which $\gamma^0\Gamma^0$
is diagonal,
it is straightforward to verify 
that the expansion (\ref{aexpansion})
indeed converges and is consistent
with the definition of $(\cdot)^{1/2}$.
It can therefore be used to determine
$A$ to arbitrary order.

\section{Symmetries of the Dispersion Relation}
\label{disp}

An explicit expression
for single-particle dispersion relation \rf{disprel}
can be found by expanding the determinant:
\cite{kl01}
\begin{eqnarray}
\det(\Gamma^\mu{\lambda}_{\mu}-M)&
=4(V_{[\mu}A_{\nu]}
-V_{\mu}V_{\nu}+A_{\mu}A_{\nu}
+PT_{\mu\nu}-S\tilde{T}_{\mu\nu}
+T_{\mu\alpha}T^{\alpha}_{\;\;\,\nu}
+\tilde{T}_{\mu\alpha}
\tilde{T}^{\alpha}_{\;\;\,\nu})^2
&
\nonumber\\
&+(V^2-A^2-S^2-P^2)^2-4(V^2-A^2)^2
+6(\epsilon_{\mu\nu\alpha\beta}
A^{\alpha}V^{\beta})^2\; .&
\label{convenientform}
\end{eqnarray}
Here,
$S$, $P$, $V^{\mu}$, $A^{\mu}$, and
$T^{\mu\nu}$ are given in
$\lambda$-momentum space,
and $V_{[\mu}A_{\nu]}=V_{\mu}A_{\nu}
-A_{\mu}V_{\nu}$
denotes the antisymmetric part.
Relation \rf{convenientform}
also follows directly from Eqs.\ \rf{cofdef}
and \rf{cofdef2}.
The position-space version of
Eq.\ \rf{convenientform}
can be used to cast Eq.\ \rf{fourth}
into a more explicit form.

Further insight
about the structure
of dispersion relation (\ref{disprel})
can be gained
by analysing its properties
under charge conjugation C,
parity inversion P, and time reversal T.
For example, 
in the absence of explicit expressions for the eigenenergies, 
these investigations
can be used to obtain information
about their degeneracy.
One such transformation,
charge conjugation C,
has been considered
previously. \cite{kl01}
It was shown
that the two sets of roots
\begin{eqnarray}
\left\{
\lambda^0_{-(\alpha)}(\vec{\lambda},m,
a^{\mu}, b^{\mu}, c^{\mu \nu}, d^{\mu \nu},
e^{\mu}, f^{\mu}, g^{\mu \nu \rho},
H^{\mu \nu})\right\}
=\left\{-\lambda^0_{+(\beta)}(-\vec{\lambda},m,
-a^{\mu}, b^{\mu}, c^{\mu \nu}, -d^{\mu \nu},
-e^{\mu}, -f^{\mu}, g^{\mu \nu \rho},
-H^{\mu \nu})\right\}
\label{csymmetry}
\end{eqnarray}
are identical.
In this section,
we employ the same methodology
to find additional symmetries
of the dispersion relation (\ref{disprel})
and point out some subtleties
regarding the labeling of the roots.
 
The idea is as follows.
Multiplication of the modified
Dirac operator
$(\Gamma^\mu{\lambda}_{\mu}-M)$
with a nonsingular,
$\lambda$-independent matrix
contributes only
a nonzero multiplicative factor
to the determinant in Eq.\ (\ref{disprel}) 
leaving this equation, 
and thus its roots, 
unchanged.
The determinant remains also invariant under
transposition or complex conjugation
of the modified Dirac operator.
The latter is true because
$\det(\Gamma^\mu{\lambda}_{\mu}-M)$
is real,
which follows
from Eq.\ (\ref{convenientform})
and Def.\ (\ref{lagr2ex}).

We first consider
\beq
(\Gamma^\mu{\lambda}_{\mu}-M)
\quad \rightarrow \quad
\gamma^0 (\Gamma^\mu{\lambda}_{\mu}-M)
\gamma^0
\label{P1}
\; ,
\eeq
which corresponds to parity inversion P
in spinor space.
We remark in passing that
$\:\det(\Gamma^\mu{\lambda}_{\mu}-M)\:$
remains unchanged,
since the overall factor induced
is $\det(\gamma^0\gamma^0)=1$.
It follows that
the dispersion relation (\ref{disprel})
is invariant under
\begin{eqnarray}
\left\{\lambda^{\mu}, m,
a^{\mu}, b^{\mu}, c^{\mu \nu}, d^{\mu \nu},
e^{\mu}, f^{\mu}, g^{\mu \nu \rho},
H^{\mu \nu}\right\}\quad\rightarrow
\quad\left\{\lambda_{\mu}, m,
a_{\mu}, -b_{\mu}, c_{\mu \nu}, -d_{\mu \nu},
e_{\mu}, -f_{\mu}, g_{\mu \nu \rho},
H_{\mu \nu}\right\}
\label{P2}
\; .
\end{eqnarray}
Hence, the two sets
\begin{eqnarray}
\left\{\lambda^0_{+ (\alpha)}(\vec{\lambda},m
,a^{\mu}, b^{\mu}, c^{\mu \nu}, d^{\mu \nu},
e^{\mu}, f^{\mu}, g^{\mu \nu \rho},
H^{\mu \nu})\right\}=
\left\{\lambda^0_{+ (\beta)}(-\vec{\lambda},m
,a_{\mu}, -b_{\mu}, c_{\mu \nu}, -d_{\mu \nu},
e_{\mu}, -f_{\mu}, g_{\mu \nu \rho},
H_{\mu \nu})\right\}
\label{P3}
\; ,
\end{eqnarray}
each containing 
the two positive-valued solutions
of Eq.\ (\ref{disprel}),
must be identical.
The result
for the remaining sets
of the two negative-valued roots
is obtained
by replacing the subscript $+$
by $-$ in Eq.\ \rf{P3}.

Next, we investigate
spinor-space time reversal
given by
\beq
(\Gamma^\mu{\lambda}_{\mu}-M)
\quad \rightarrow \quad
i\gamma^5 C
(\Gamma^\mu{\lambda}_{\mu}-M)^*
iC\gamma^5
\label{T1}
\; ,
\eeq
where $*$ denotes complex conjugation
and $C$ is the usual charge-conjugation matrix.
Again, 
$\:\det(\Gamma^\mu{\lambda}_{\mu}-M)\:$
is left unchanged.
The resulting symmetry between
the positive-valued roots
of the dispersion relation (\ref{disprel})
takes the form
\begin{eqnarray}
\left\{\lambda^0_{+ (\alpha)}(\vec{\lambda},m
,a^{\mu}, b^{\mu}, c^{\mu \nu}, d^{\mu \nu},
e^{\mu}, f^{\mu}, g^{\mu \nu \rho},
H^{\mu \nu})\right\}=
\left\{\lambda^0_{+ (\beta)}(-\vec{\lambda},m
,a_{\mu}, b_{\mu}, c_{\mu \nu}, d_{\mu \nu},
e_{\mu}, -f_{\mu}, -g_{\mu \nu \rho},
-H_{\mu \nu})\right\}
\label{T3}
\; .
\end{eqnarray}
For the corresponding expression
involving the negative-valued solutions,
the subscript $+$ must be
changed to $-$.

Note that the above arguments,
which generated
\rf{csymmetry}, \rf{P3} and \rf{T3},
provide only equalities
between {\it sets} of roots.
In order to find relations
between the individual roots,
additional considerations are needed.
Charge conjugation on one hand,
and parity inversion and time reversal
on the other hand
have to be treated separately.

We begin by discussing charge conjugation.
Equality \rf{csymmetry}
relates positive- and negative-valued
eigenenergies.
It therefore follows already at this point
that in principle relation \rf{csymmetry}
and the knowledge of one root
suffices to construct another solution.
The remaining less important question is
how the subscript $(\alpha)$
behaves under C.
If an additional conserved quantity
commuting with $H$ is known
(such as a spin component or helicity
in the conventional case),
the subscript $(\alpha)$ may be used
to label its eigenvalues.
A definite correspondence
can then be determined by investigating
the behavior of this conserved quantity
under C.
In the present case without the knowledge
of such an additional conserved quantity,
the label $(\alpha)$ becomes essentially arbitrary
and can therefore be chosen freely.\cite{note2} 
Our conventions agree with the conventional ones
in the following sense:\cite{note1}
the labels of eigenvalues and eigenspinors
match and change under charge conjugation.
This produces after reinterpretation
\begin{eqnarray}
E^{(1,2)}_v(\vec{p},m,
a^{\mu}, b^{\mu}, c^{\mu \nu}, d^{\mu \nu},
e^{\mu}, f^{\mu}, g^{\mu \nu \rho},
H^{\mu \nu})
=E^{(2,1)}_u(\vec{p},m,
-a^{\mu}, b^{\mu}, c^{\mu \nu}, -d^{\mu \nu},
-e^{\mu}, -f^{\mu}, g^{\mu \nu \rho},
-H^{\mu \nu})
\; ,
\label{csymmetry2}
\end{eqnarray}
where $E^{(1,2)}_{u,v}$ denotes
the zero-components of
$p^{(1,2)}_{u,v}$ 
defined by (\ref{reinterpr1}).
We remark that this labeling agrees
with our previous choice.\cite{kl01}

We now turn to parity inversion
and time reversal.
The equalities (\ref{P3}) and (\ref{T3})
provide a correspondence between
roots of the same sign.
As opposed to the previous case for C,
it is therefore unclear {\it a priori}
whether P and T individually can be used
to construct additional eigenvalues
from a known one.
However, the combined transformation PT
should give a different root:
it connects different eigenspinors,\cite{note3}
and according to our above conventions,
this fact should be reflected in the labels of the eigenenergies.
The invariance of the dispersion relation
(\ref{disprel}) under the transformation PT
yields with this choice of labeling after reinterpretation
\beq
E^{(1,2)}_{u,v}(\vec{p},m
,a^{\mu}, b^{\mu}, c^{\mu \nu}, d^{\mu \nu},
e^{\mu}, f^{\mu}, g^{\mu \nu \rho},
H^{\mu \nu})=E^{(2,1)}_{u,v}(\vec{p},m
,a^{\mu}, -b^{\mu}, c^{\mu \nu},
-d^{\mu \nu},
e^{\mu}, f^{\mu}, -g^{\mu \nu \rho},
-H^{\mu \nu})
\; .
\label{PT3}
\eeq
To determine equalities between the individual roots in \rf{P3} and \rf{T3},
a definite labeling scheme is needed.
Without knowledge of an additional conserved operator
this becomes a matter of choice
constrained only by Eq.\ (\ref{PT3}).
Contrary to the charge-conjugation case,
there is more freedom
even at the conventional level:
the two customary labels 
spin projection onto a fixed direction
or helicity
behave differently under both P and T.
For definiteness,
we choose the label $(\alpha)$ to change under P,
but not under T.
This agrees with the behavior of the conventional helicity labeling.

We mention that from any other combination of C, P, and T
additional correspondences can be constructed straightforwardly.
However, if one eigenenergy
(with functional dependence agreeing with our choice of labeling)
is known explicitly,
the symmetries (\ref{csymmetry2}) and (\ref{PT3}) suffice
to determine the remaining three.
As an illustrative example,
consider the following case:
all Lorentz-violating parameters except $a_{\mu}$ and $b_0$ are zero.
The eigenenergies are then given by \cite{ck97}
\bea
E^{(\alpha)}_u=\sqrt{m^2+\left(|\vec{p}-\vec{a}|+(-1)^{\alpha}b_0\right)^2}
+a_0\; ,
\quad\;
E^{(\alpha)}_v=\sqrt{m^2+\left(|\vec{p}+\vec{a}|-(-1)^{\alpha}b_0\right)^2}
-a_0\; .
\label{abcase}
\eea 
Suppose only one of the above energies, say $E^{(1)}_u
=[m^2+(|\vec{p}-\vec{a}|-b_0)^2]^{1/2}
+a_0$, is known.
According to Eq.\ (\ref{PT3}), the eigenvalue $E^{(2)}_u$
for the other positive-energy solution can then be obtained by keeping
the sign of $a_{\mu}$ the same, but reversing the sign of $b_0$
in complete agreement with Eq.\ (\ref{abcase}).
Similarly, the symmetry (\ref{csymmetry2}) permits
the determination of the antiparticle energy $E^{(2)}_v$
by keeping all signs unchanged except for the one of $a_{\mu}$,
which has to be reversed.
Again, the result is identical to the one given by (\ref{abcase}).
The remaining antiparticle energy $E^{(1)}_v$
can be found in the same way as $E^{(2)}_v$,
but starting from $E^{(2)}_u$ instead of $E^{(1)}_u$.

The above method for the construction
of additional eigenvalues from a known root
has to be taken with a grain of salt.
It is required
that the functional dependence of the given eigenenergy
on the Lorentz-violating parameters
is consistent with our above choice of labeling.
The following example illustrates this issue.
Consider a model in which Lorentz breaking
can be described by $a^{\mu}$ and $\vec{b}$ only.
The following expressions for the reinterpreted eigenenergies
satisfy its dispersion relation: \cite{ck97}
\bea
E^{\pm}_u=\left[m^2+(\vec{p}-\vec{a})^2\pm2\sqrt{m^2{\vec{b}}^{2}+(\vec{b}
\cdot{\left(\vec{p}-\vec{a})\right)}^2}+{\vec{b}}^{2}\right]^{1/2}+a_0
\; , \nonumber\\
E^{\pm}_v=\left[m^2+(\vec{p}+\vec{a})^2\pm2\sqrt{m^2{\vec{b}}^{2}+(\vec{b}
\cdot{\left(\vec{p}+\vec{a})\right)}^2}+{\vec{b}}^{2}\right]^{1/2}-a_0
\; .
\label{bvec}
\eea
Suppose again that only one of these eigenenergies, say $E^{+}_u$,
were known.
Employing symmetry (\ref{PT3}), i.e., reversing the sign of $\vec{b}$,
does not yield any of the other eigenenergies.
This can be traced to the fact
that the functional dependence in Eq.\ (\ref{bvec})
does not agree with our convention that the labels should change
under PT.
In Eq.\ (\ref{bvec}), 
the labels reflect the sign of a square root
in the expression rather than being related to the PT transformation.
We remark that a suitable labeling can be obtained by multiplying
the inner square roots in Eq.\ (\ref{bvec}) by a quantity $D$
that can take the values $+1$ and $-1$ and changes sign
when $\vec{b}$ is reversed
(i.e., a possible choice would be
$D=\vec{b}\cdot\vec{B}/|\vec{b}\cdot\vec{B}|$,
where $\vec{B}$ is arbitrary but fixed and satisfies
$\vec{b}\cdot\vec{B}\neq0$).

The symmetries (\ref{csymmetry2}) and (\ref{PT3})
can also be used to find parameter combinations
yielding degenerate roots.
It follows from Eq.\ (\ref{PT3}) that for
\beq
b^{\mu}=d^{\mu \nu}=g^{\mu \nu \rho}=H^{\mu \nu}=0
\; ,
\label{degen1}
\eeq
roots of the same sign are equal.
Thus, Eq. (\ref{degen1}) provides a sufficient condition
for energy equality of two (distinct) particle states
of a given 3-momentum.
The same holds true for the antiparticles.
Suppose the Lorentz-violating parameters obey
\beq
a^{\mu}=d^{\mu \nu}=e^{\mu}=f^{\mu}=H^{\mu \nu}=0
\; ,
\label{degen2}
\eeq
or
\beq
a^{\mu}=b^{\mu}=e^{\mu}=f^{\mu}=g^{\mu \nu \rho}=0
\; .
\label{degen3}
\eeq
Either one of the conditions (\ref{degen2}) and (\ref{degen3}) is sufficient
for an energy degeneracy
such that for each particle
there exists an antiparticle of equal 4-momentum.
This can be verified 
by using the symmetries (\ref{csymmetry2}) and (\ref{PT3}).
As a corollary of the above discussion 
we remark
that if $c^{\mu \nu}$ is the only non-vanishing Lorentz-violating parameter,
then all four eigenenergies
become degenerate after reinterpretation.

\section{First-Order Approximation of the Eigenenergies}
\label{1eigen}

In principle,
the dispersion relation \rf{convenientform} and Eq.\ \rf{lagr2ex}
allow the determination
of the exact eigenenergies
at a given 3-momentum
in the presence of Lorentz violation.
In many circumstances,
however,
only leading-order corrections
to the conventional eigenenergies
are of interest.
They can be obtained with the method described below.

With the aid of generalized Foldy--Wouthuysen techniques
one can find a (momentum-dependent) unitary matrix $U$
transforming the Hamiltonian (\ref{hamiltonian})
into the following block-diagonal form: \cite{kl99}
\beq
U^{\dagger}HU=\left(
\begin{array}{cc}
h_{\rm rel}& 0 \\
0 & \overline{h}_{\rm rel}
\end{array}
\right)
\; ,
\label{hdiag}
\eeq
where the two 2$\times$2 matrices $h_{\rm rel}$ and $\overline{h}_{\rm rel}$
are the respective Hamiltonians for the fermion and the antifermion.
From the procedure it is obvious
that the eigenvalues of the matrices $h_{\rm rel}$ and $\overline{h}_{\rm rel}$
are the respective particle and antiparticle energies.
To make further progress, consider the expansion
\beq
h_{\rm rel}=h_0+\sum_{j=1}^3 h_j \sigma_j
\label{hrel}
\eeq
of $h_{\rm rel}$
with respect to the basis $\{I,\sigma_j\}$.
Here, $I$ is the 2$\times$2 identity
and $\sigma_j$ are the usual Pauli matrices.
The components $h_0$, \ldots, $h_3$ depend on the 3-momentum,
the mass, and the parameters for Lorentz breaking
as determined by the Foldy--Wouthuysen transformation.
They yield the fermion eigenenergies by means of the equation
\beq
E^{(1,2)}_{u}=h_0\pm \sqrt{\sum_{j=1}^3 h_j h_j}
\; .
\label{eigen}
\eeq
Note that the correspondence between
the energy superscript $(\cdot)$ and the sign of the square root
in (\ref{eigen}) is only constrained by the symmetry (\ref{PT3}).

This method is suitable for extracting the leading-order approximation
of the eigenenergies because the components $h_0$, \ldots, $h_3$
are known to first order in the Lorentz-violating parameters: \cite{kl99}
\bea
h_0
&=& \ga m 
 + (a_0-\fr{mc_{00}}{\ga}-me_0)
     +\Big[
       a_j-\ga m(c_{0j}+c_{j0})-me_j       
     \Big] \fr{p^j}{\ga m}
     -(c_{jk}-\et_{jk}c_{00})\fr{p^j p^k}{\ga m}\; ,
 \nonumber \\
h_j&=&
       -\fr{1}{\ga}b_j+md_{j0}+\half{\ve^{kl}}_{j}H_{kl}
       -\fr{1}{2\ga}m{\ve^{kl}}_{j}g_{kl0}
 \nonumber \\
&& +\Big[
    \et_{jk}b_{0}+m(d_{jk}-\et_{jk}d_{00})+{\ve^{l}}_{kj}H_{0l}
    -\ga m{\ve^{lm}}_{j}(\half g_{lmk}-\et_{km}g_{l00})
   \Big] \fr{p^k}{\ga m}
 \nonumber \\
&& +\Big[
     \fr{(\ga-1)m}{{\vec{p}}^{\:2}}
     (b_k+md_{k0}+\half {\ve^{mn}}_{k}H_{mn}
       +\half m{\ve^{mn}}_{k}g_{mn0})
     \et_{jl}
    -(d_{0k}+d_{k0})\et_{jl}
    +{\ve^{m}}_{lj}(g_{m0k}+g_{mk0})
   \Big] \fr{p^k p^l}{\ga m}
 \nonumber \\
&& +\fr{(\ga-1)}{{\vec{p}}^{\:2}} 
   \Big[
    -(d_{kl}-\et_{kl}d_{00}) 
    -\half {\ve^{nq}}_{l}g_{nqk}
   \Big] \et_{jm} \fr{p^k p^l p^m}{\ga }
\; ,
\label{FW_ul}
\eea
where $\ga \equiv \sqrt{1+{\vec{p}}^{\:2}/m^2}$
is the usual relativistic gamma factor,
and the totally antisymmetric rotation tensor
$\ve^{jkl}$ satisfies $\ve_{123} = +1$ and  
$\ve^{jkl} = - \ve_{jkl}$.
Note that the parameter $f^{\mu}$ does not contribute to the eigenenergies
at leading order.
We remark that the symmetries
(\ref{csymmetry2}) and (\ref{PT3})
permit the construction of the antifermion energies.

As an illustration, 
consider the previously considered $(a_{\mu},b_0)$ model
with the eigenenergies (\ref{abcase}).
For this model, we have $h_0=\gamma m+a_0-\vec{p}\cdot\vec{a}/\gamma m$
and $h_j=b_0p_j/\gamma m$.
For this case, Eq.\ (\ref{eigen}) yields
\beq
E^{(1,2)}_u=\sqrt{m^2+{\vec{p}}^{\:2}}+a_0
-\fr{\vec{p}\cdot\vec{a}\pm b_0|\vec{p}|}{\sqrt{m^2+{\vec{p}}^{\:2}}}
\; .
\label{1abcase}
\eeq
One can verify 
that Eqs.\ (\ref{abcase}) and (\ref{1abcase})
agree to leading order in the Lorentz-violating coefficients 
provided the upper and lower signs in Eq.\ (\ref{1abcase}) 
are identified with the labels
$\alpha=1$ and $\alpha=2$, respectively.
The corresponding antiparticle energies to first order
can now be obtained with the aid of symmetry (\ref{csymmetry2}),
as discussed previously.

\section{Symmetries of the Eigenspinors}
\label{spisym}

It is necessary to begin this section
by introducing our conventions and some more notation.
The four eigenspinors
$w^{(\alpha)}_{\pm}(\vec{\lambda})$
of $H(\vec{\lambda})$
are determined by
\beq
\left(H(\vec{\lambda})
-\lambda^0_{\pm (\alpha)}\right)
w^{(\alpha)}_{\pm}(\vec{\lambda})=0
\; .
\label{eigham}
\eeq
We have used that
the dispersion relation (\ref{disprel}),
and thus its roots
$\lambda^0_{\pm (\alpha)}$,
remain unchanged
under the field redefinition.
The eigenspinors
$w^{(\alpha)}_{\pm}(\vec{\lambda})$
are related
to the observer-covariant
momentum-space spinors
$W^{(\alpha)}_{\pm}(\vec{\lambda})$
obeying Eq.\ (\ref{spinordet}) by
\beq
W^{(\alpha)}_{\pm}(\vec{\lambda})=
Aw^{(\alpha)}_{\pm}(\vec{\lambda})
\; ,
\label{relation}
\eeq
where $A$ is the field-redefinition matrix
discussed earlier.
After reinterpretation
of the negative-energy solutions
we denote the eigenspinors
of $H$ by $u^{(\alpha)}(\vec{p})$
and $v^{(\alpha)}(\vec{p})$
in complete analogy to
Def.\ (\ref{reinterpr2}).
Thus, 
the transformation (\ref{relation})
remains valid
even after the reinterpretation.
The four eigenspinors
for a given momentum,
which can be taken as orthogonal,
span spinor space.
Our choice of normalization is
\beq
u^{(\alpha)\dagger}(\vec{p})
u^{(\alpha^{\prime})}(\vec{p})=
\delta^{\alpha\alpha^{\prime}}
\fr{E^{(\alpha)}_u}{m}
\; ,\quad\;
v^{(\alpha)\dagger}(\vec{p})
v^{(\alpha^{\prime})}(\vec{p})=
\delta^{\alpha\alpha^{\prime}}
\fr{E^{(\alpha)}_v}{m}
\; ,\quad\;
u^{(\alpha)\dagger}(\vec{p})
v^{(\alpha^{\prime})}(-\vec{p})
= 0
\label{normalize}
\; .
\eeq
Note that the physical
spinors
$w^{(\alpha)}_{\pm}$,
and thus $u^{(\alpha)}$ and $v^{(\alpha)}$,
fail to be observer Lorentz covariant
due to the frame dependence of $A$.

The discrete transformations C, P, and T
determine correspondences between sets of spinors,
paralleling the eigenenergy case.
For the charge-conjugation transformation
our previous result \cite{kl01}
\beq
\left\{W^{(\alpha)}_{-}
(\vec{\lambda},m,
a^{\mu}, b^{\mu}, c^{\mu \nu}, d^{\mu \nu},
e^{\mu}, f^{\mu}, g^{\mu \nu \rho},
H^{\mu \nu})\right\}
\propto\left\{{W^{(\beta)}_{+}}^{c}(-\vec{\lambda},m,
-a^{\mu}, b^{\mu}, c^{\mu \nu}, -d^{\mu \nu},
-e^{\mu}, -f^{\mu}, g^{\mu \nu \rho},
-H^{\mu \nu})\right\}
\label{cspinor1}
\eeq
holds,
which we provide here for completeness.
The charge-conjugated spinor
$W^{c}\equiv C\overline{W}^{T}$
is defined 
with the conventional charge-conjugation matrix $C$.
A $\propto$ sign,
such as the one in relation (\ref{cspinor1}), 
is to be understood as follows. 
For each each spinor in one set
there exists a spinor in the other set such that
the two spinors are linearly dependent.

The parity transformation (\ref{P1}) induces the following
relation
\beq
\left\{W^{(\alpha)}_{+}
(\vec{\lambda},m,
a^{\mu}, b^{\mu}, c^{\mu \nu}, d^{\mu \nu},
e^{\mu}, f^{\mu}, g^{\mu \nu \rho},
H^{\mu \nu})\right\}
\propto\left\{{W^{(\beta)}_{+}}^{p}
(-\vec{\lambda},m
,a_{\mu}, -b_{\mu}, c_{\mu \nu}, -d_{\mu \nu},
e_{\mu}, -f_{\mu}, g_{\mu \nu \rho},
H_{\mu \nu})
\right\}
\; ,
\label{pspinor1}
\eeq
where $W^{p}\equiv \gamma^{0}W$ denotes the parity-inverted spinor as usual.
The relation for the negative-eigenvalue spinors is obtained
by replacing the subscripts $+$ with $-$.
The result for time reversal (\ref{T1}) is given by
\beq
\left\{W^{(\alpha)}_{+}
(\vec{\lambda},m,
a^{\mu}, b^{\mu}, c^{\mu \nu}, d^{\mu \nu},
e^{\mu}, f^{\mu}, g^{\mu \nu \rho},
H^{\mu \nu})\right\}
\propto\left\{{W^{(\beta)}_{+}}^{t}
(-\vec{\lambda},m
,a_{\mu}, b_{\mu}, c_{\mu \nu}, d_{\mu \nu},
e_{\mu}, -f_{\mu}, -g_{\mu \nu \rho},
-H_{\mu \nu})
\right\}
\; .
\label{tspinor1}
\eeq
Here, $W^{t}\equiv -i\gamma^{5}CW^{*}$
is the conventional time-reversed spinor.
Again, changing the subscripts from $+$ to $-$ yields the corresponding
relation for the remaining eigenspinors.

To find correspondences between individual eigenspinors,
a definite labeling scheme for the spinors must be selected.
The associated subtleties are analogous to the eigenenergy case
and do not require additional discussion.
Our previous convention 
that the labeling of the roots and eigenspinors matches
leads after reinterpretation to the symmetries
\beq
V^{(1,2)}
(\vec{p},m,
a^{\mu}, b^{\mu}, c^{\mu \nu}, d^{\mu \nu},
e^{\mu}, f^{\mu}, g^{\mu \nu \rho},
H^{\mu \nu})
=\zeta{U^{(2,1)}}^{c}(\vec{p},m,
-a^{\mu}, b^{\mu}, c^{\mu \nu}, -d^{\mu \nu},
-e^{\mu}, -f^{\mu}, g^{\mu \nu \rho},
-H^{\mu \nu})
\label{cspinor2}
\eeq
and
\begin{eqnarray}
U^{(1)}
(\vec{p},m,
a^{\mu}, b^{\mu}, c^{\mu \nu}, d^{\mu \nu},
e^{\mu}, f^{\mu}, g^{\mu \nu \rho},
H^{\mu \nu})
& = & \eta{U^{(2)}}^{pt}(\vec{p},m,
a^{\mu}, -b^{\mu}, c^{\mu \nu}, -d^{\mu \nu},
e^{\mu}, f^{\mu}, -g^{\mu \nu \rho},
-H^{\mu \nu})
\label{ptspinor2}
\end{eqnarray}
resulting from C and PT, respectively.
Here, 
$\zeta$ and $\eta$ are (possibly spinor-dependent) proportionality factors,
and the superscript $pt$ stands for the combined spinor transformations
P and T defined above.
The relation arising from PT,
but involving the spinors $V^{(1,2)}$
can be obtained by replacing $U$ with $V$ in Eq.\ (\ref{ptspinor2}).
If one eigenspinor is known explicitly
(with functional dependence agreeing with our choice of labeling),
the symmetries (\ref{cspinor2}) and (\ref{ptspinor2}) can in principal be used
to construct the remaining three.

As an illustration, 
we again consider the $(a_{\mu},b_0)$ model. 
Its eigenspinors in Pauli-Dirac representation
are \cite{ck97}
\bea
U^{(\alpha)}(\vec{p},m,a^{\mu},b^0)=\left(\fr{E^{(\alpha)}_u
(E^{(\alpha)}_u-a_0+m)}{2m(E^{(\alpha)}_u-a_0)}\right)^{1/2}
\left(
\begin{array}{c}
\phi^{(\alpha)}(\vec{p}-\vec{a})\vspace{.2cm}\\
\frac{\displaystyle{-(-1)^{\alpha}|\vec{p}-\vec{a}|-b_0}}
{\displaystyle{E^{(\alpha)}_u-a_0+m}}
\phi^{(\alpha)}(\vec{p}-\vec{a})
\end{array}
\right)\; ,\nonumber\\
\nonumber\\
V^{(\alpha)}(\vec{p},m,a^{\mu},b^0)=\left(\fr{E^{(\alpha)}_v
(E^{(\alpha)}_v+a_0+m)}{2m(E^{(\alpha)}_v+a_0)}\right)^{1/2}
\left(
\begin{array}{c}
\frac{\displaystyle{-(-1)^{\alpha}|\vec{p}+\vec{a}|+b_0}}
{\displaystyle{E^{(\alpha)}_v+a_0+m}}
\phi^{(\alpha)}(\vec{p}+\vec{a})\vspace{.2cm}\\
\phi^{(\alpha)}(\vec{p}+\vec{a})
\end{array}
\right)\; ,
\label{abspinors}
\eea
where the two-component spinors $\phi^{(\alpha)}(\vec{k})$
are given by
\beq
\phi^{(1)}(\vec{k})=
\left(
\begin{array}{c}
\cos (\theta/2)\\
e^{i\varphi}\sin (\theta/2)
\end{array}
\right)
\; ,
\quad\;
\phi^{(2)}(\vec{k})=
\left(
\begin{array}{c}
-e^{-i\varphi}\sin (\theta/2)\\
\cos (\theta/2)
\end{array}
\right)
\; .
\label{2spinors}
\eeq
Here, $\theta$ and $\varphi$ are the spherical-polar angles
subtended by $\vec{k}$.
Suppose that only one of the spinors in Eq.\ \rf{abspinors},
say $U^{(1)}(\vec{p},m,a^{\mu},b^0)$, is known.
With the symmetry (\ref{ptspinor2}) 
one can now construct $U^{(2)}(\vec{p},m,a^{\mu},b^0)$ up to a constant:
in $U^{(1)}(\vec{p},m,a^{\mu},b^0)$,
the sign of $b^0$ has to be reversed.
Note that this entails changing $E^{(1)}_u$ to $E^{(2)}_u$
by virtue of Eq.\ (\ref{PT3}).
Complex conjugation resulting from time reversal
affects only $\phi^{(1)}(\vec{p}-\vec{a})$
because all other quantities in the expression
are real for this specific model.
The final step is multiplication with the matrix
$-i\gamma^3\gamma^1\gamma^0=\left(\begin{array}{cc}
-\sigma^2 & 0\\0 & \sigma^2\end{array}\right)$,
where $\sigma^2$ denotes the usual Pauli matrix
associated with 2-direction.
This is somewhat simplified
by observing that $\sigma^2{\phi^{(1)}}^*(\vec{k})=i\phi^{(2)}(\vec{k})$.
Comparison with Eq.\ (\ref{abspinors}) shows
that the resulting spinor 
is indeed $U^{(2)}(\vec{p},m,a^{\mu},b^0)$ 
up to a factor of $-i$.
The symmetry (\ref{cspinor2}) determines (up to constants)
the remaining two spinors $V^{(1)}(\vec{p},m,a^{\mu},b^0)$ and
$V^{(2)}(\vec{p},m,a^{\mu},b^0)$
from $U^{(2)}(\vec{p},m,a^{\mu},b^0)$
and $U^{(1)}(\vec{p},m,a^{\mu},b^0)$, respectively:
reverse the sign of $a^{\mu}$, complex conjugate and multiply by
$i\gamma^2$.
This procedure yields $V^{(2)}(\vec{p},m,a^{\mu},b^0)$ exactly
and $V^{(1)}(\vec{p},m,a^{\mu},b^0)$ up to a relative minus sign.

With an explicit labeling scheme 
like that selected for the eigenenergies,  
additional relations between the spinors can be determined
using other combinations
of Eqs.\ (\ref{cspinor1}), (\ref{pspinor1}), and (\ref{tspinor1}).
The corresponding symmetries for the physical spinors defined earlier
can be obtained
by replacing $U^{(1,2)}$ and $V^{(1,2)}$
in Eqs.\ (\ref{cspinor2}) and (\ref{ptspinor2})
by $u^{(1,2)}$ and $v^{(1,2)}$,
respectively.
Note that the field-redefinition matrix $A$
depends on the Lorentz-violating parameters.
For example, to construct $u^{(\alpha)}(\vec{p},m,
-a^{\mu}, b^{\mu}, c^{\mu \nu}, -d^{\mu \nu},
-e^{\mu}, -f^{\mu}, g^{\mu \nu \rho},
-H^{\mu \nu})$ from $U^{(\alpha)}(\vec{p},m,
a^{\mu}, b^{\mu}, c^{\mu \nu}, d^{\mu \nu},
e^{\mu}, f^{\mu}, g^{\mu \nu \rho},
H^{\mu \nu})$ and $A(c^{\mu 0}, d^{\mu 0},
e^{0}, f^{0}, g^{\mu \nu 0})$,
the appropriate sign changes have to implemented {\it both}
in $U^{(\alpha)}$ and in $A$.

\section{Generalization of the Conventional Spinor Projectors}
\label{spipro}

In the ordinary Dirac case,
the spinor matrices 
that project on the positive- and negative-energy eigenspaces
\beq
\pm\sum_{\alpha=1}^2 w^{(\alpha)}_{\pm}
\otimes\overline{w}^{(\alpha)}_{\pm}
=\fr{\lambda_{\pm}\hspace{-.39cm}/\hspace{.21cm}+m}{2m}
\label{conpro}
\eeq
are an indispensable tool in numerous calculations. 
To obtain the Lorentz-violating analog,
we fix an arbitrary 3-momentum
$\vec{\lambda}$
and express
the left-hand side of Eq.\ (\ref{cofdef})
in terms
of the Hamiltonian (\ref{hamiltonian}):
\beq
{\rm cof}(\Gamma_{\mu}\lambda^{\mu}-M)
=\det (\Gamma^0)\prod_{[j]}
(\lambda^0-\lambda^0_{[j]})
(\gamma^0\Gamma^0)^{-1/2}
(\lambda^0-H)^{-1}
(\gamma^0\Gamma^0)^{-1/2}
\gamma^0
\; .
\label{cofham}
\eeq
Here and in what follows,
the dependence of the eigenvalues,
the eigenspinors, and the Hamiltonian
on the fixed $\vec{\lambda}$
is omitted for brevity.
The two positive and two negative 
eigenvalues $\lambda^0_{+(1,2)}$ and
$\lambda^0_{-(1,2)}$
of $H$ are denoted collectively by 
$\lambda^0_{[j]}$,
where $[j]\in\{-(2),-(1),+(1),+(2)\}$.
The product in Eq.\ (\ref{cofham})
runs over all four of these eigenvalues.
Since the Hamiltonian (\ref{hamiltonian})
is hermitian,
there exists a spinor basis
in which $H$ is diagonal.
In this basis,
we have explicitly
\beq
{\rm cof}
(\Gamma_{\mu}\lambda^{\mu}-M)=
\det (\Gamma^0)
(\gamma^0\Gamma^0)^{-1/2}
P_{\lambda^0}
(\gamma^0\Gamma^0)^{-1/2}
\gamma^0
\; ,
\label{cofex}
\eeq
where the diagonal matrix $P_{\lambda^0}$
is given by
\beq
P_{\lambda^0}\equiv
\left(
\begin{array}{l}
(\lambda^0-\lambda^0_{+(2)})
(\lambda^0-\lambda^0_{-(1)})
(\lambda^0-\lambda^0_{-(2)})
\lefteqn{
\hspace{6.4cm}\cdots\quad 0}\\
\hspace{2.5cm}(\lambda^0-\lambda^0_{+(1)})
(\lambda^0-\lambda^0_{-(1)})
(\lambda^0-\lambda^0_{-(2)})
\lefteqn{\hspace{4.76cm}\vdots}\\
\lefteqn{\hspace{.04cm}\vdots}
\hspace{5cm}(\lambda^0-\lambda^0_{+(1)})
(\lambda^0-\lambda^0_{+(2)})
(\lambda^0-\lambda^0_{-(2)})
\vspace{.25cm}\\
\lefteqn{0\quad\cdots}
\hspace{7.5cm}(\lambda^0-\lambda^0_{+(1)})
(\lambda^0-\lambda^0_{+(2)})
(\lambda^0-\lambda^0_{-(1)})
\end{array}
\right)
\; .
\label{upsilon}
\eeq
For
$\lambda^0\rightarrow\lambda^0_{+(1)}$, 
the nondegenerate case
$\lambda^0_{+(1)}\neq\lambda^0_{+(2)}$
and the degenerate case
$\lambda^0_{+(1)}=\lambda^0_{+(2)}$ 
have to be distinguished.

In the nondegenerate case,
the matrix $P_{\lambda^0}$ simplifies to
\beq
P_{\lambda^0_{+(1)}}=
\left(
\begin{array}{cccc}
(\lambda^0_{+(1)}-\lambda^0_{+(2)})
(\lambda^0_{+(1)}-\lambda^0_{-(1)})
(\lambda^0_{+(1)}-\lambda^0_{-(2)})
& 0 & 0 & 0\\
0 & 0 & 0 & 0\\
0 & 0 & 0 & 0\\
0 & 0 & 0 & 0
\end{array}
\right)
\; ,
\label{nupsilon}
\eeq
which can be written
$P_{\lambda^0_{+(1)}}=
(H-\lambda^0_{+(2)})(H-\lambda^0_{-(1)})
(H-\lambda^0_{-(2)})$. 
Thus, $P_{\lambda^0_{+(1)}}$
is proportional to the projector
on the $\lambda^0_{+(1)}$ eigenspace.
The above argument applied 
to an arbitrary nondegenerate eigenvalue $\lambda^0_{[r]}$ yields:
\beq
{\rm cof}
(\Gamma_{\mu}\lambda^{\mu}_{[r]}-M)=
\det (\Gamma^0)
(\gamma^0\Gamma^0)^{-1/2}
\prod_{[j]\neq [r]}(H-\lambda^0_{[j]})
(\gamma^0\Gamma^0)^{-1/2}
\gamma^0
\; .
\label{cofex2}
\eeq
The $P_{\lambda^0_{[r]}}$
can be expressed
in terms of the eigenspinors
in the usual way.
Our normalization (\ref{normalize})
gives explicitly
\beq
\prod_{[j]\neq[r]}
(H-\lambda^0_{[j]})=
\frac{\displaystyle{m}}
{\displaystyle{|\lambda^0_{[r]}|}}
w_{[r]}\otimes w^{\dagger}_{[r]}
\prod_{[j]\neq[r]}
(\lambda^0_{[r]}-\lambda^0_{[j]})
\label{project1}
\quad,
\eeq
where $w_{[r]}\in\{w^{(2)}_{-},w^{(1)}_{-},
w^{(1)}_{+},w^{(2)}_{+}\}$
is a shorthand notation
for the $\lambda^0_{[r]}$ eigenspinor.
One can now determine the desired projectors
for the observer-invariant eigenspinors.
We obtain for nondegenerate eigenvalues
$\lambda^0_{[r]}$:
\beq
\frac{\displaystyle{m}}
{\displaystyle{|\lambda^0_{[r]}|}}
W_{[r]}
\otimes \overline{W}_{[r]}
=\frac{\displaystyle{{\rm cof}
(\Gamma_{\mu}\lambda^{\mu}_{[r]}-M)}}
{\displaystyle{\det (\Gamma^0)
\prod_{[j]\neq[r]}^{}
(\lambda^0_{[r]}-\lambda^0_{[j]})}}
\; .
\label{projnond}
\eeq

For degenerate eigenvalues
$\lambda^0_{[q]}=\lambda^0_{[r]}$,
$\;([q]\neq[r])$,
the matrix $P_{\lambda^0_{[r]}}$
vanishes.
However,
Eq.\ (\ref{cofex}) can be modified
to
\beq
\widehat{\rm cof}
(\Gamma_{\mu}\lambda^{\mu}-M)=
\det (\Gamma^0)
(\gamma^0\Gamma^0)^{-1/2}
\widehat{P}_{\lambda^0}
(\gamma^0\Gamma^0)^{-1/2}
\gamma^0
\; ,
\label{cofex3}
\eeq
where a caret denotes
division by the appropriate
$(\lambda^0-\lambda^0_{[r]})$ factor.
The existence of $\widehat{P}_{\lambda^0}$
in the limit
$\lambda^0\rightarrow\lambda^0_{[r]}$
is immediate from Def.\ (\ref{upsilon}).
Note that the resulting matrix
$\widehat{P}_{\lambda^0_{[r]}}$
is again,
up to normalization,
the projector on the $\lambda^0_{[r]}$
eigenspace.
By virtue of Eq.\ (\ref{cofex3}),
$\widehat{\rm cof}
(\Gamma_{\mu}\lambda^{\mu}-M)$
is also well defined
for all $\lambda^0$.
Considerations similar to the ones
for a nondegenerate eigenvalue
yield the following intermediate expression
involving the eigenspinors of $H$:
\beq
\prod_{[j]\neq[q],[r]}
(H-\lambda^0_{[j]})=
\frac{\displaystyle{m}}
{\displaystyle{|\lambda^0_{[r]}|}}
\sum_{[k]=[q],[r]}
w_{[k]}\otimes w^{\dagger}_{[k]}
\prod_{[j]\neq[q],[r]}
(\lambda^0_{[r]}-\lambda^0_{[j]})
\label{project2}
\quad.
\eeq
The final result
for the case of two degenerate eigenvalues
$\lambda^0_{[q]}=\lambda^0_{[r]}$,
$\;([q]\neq[r])$
is given by
\beq
\frac{\displaystyle{m}}
{\displaystyle{|\lambda^0_{[r]}|}}
\sum_{[k]=[q],[r]}W_{[k]}
\otimes
\overline{W}_{[k]}
=\frac{\displaystyle{\widehat{\rm cof}
(\Gamma_{\mu}\lambda^{\mu}_{[r]}-M)}}
{\displaystyle{\det (\Gamma^0)
\prod_{[j]\neq[q],[r]}^{}
(\lambda^0_{[r]}-\lambda^0_{[j]})}}
\; .
\label{projd}
\eeq

For any square matrix $B$,
the relation $U^{\dagger}{\rm cof}(B)U={\rm cof}(U^{\dagger}BU)$ holds,
where $U$ is unitary.
Thus the expressions (\ref{projnond}) and (\ref{projd})
for the generalized projectors
are independent of the spinor-space basis.
We also remark that the projectors (\ref{conpro}) for the ordinary Dirac case
can be recovered from these results, as expected.
In this case,
the eigenenergies are degenerate
so that Eq.\ (\ref{projd}) must be employed.
Moreover,
no field redefinition is necessary,
so that the covariant and physical spinors are identical.
The matrix of cofactors is given by
$(\lambda^2-m^2)(\lambda\hspace{-.155cm}/+m)$,
which can be verified directly or can be obtained from Eq.\ (\ref{cofdef2}).
Assembling everything yields Eq.\ (\ref{conpro}).

As an immediate application,
the generalized projector (\ref{projnond})
permits the construction of a more explicit form of the eigenspinor
in the case when there is a nondegenerate eigenvalue $\lambda^0_{[r]}$:
\beq
W_{[r]}(\vec{\lambda})=N_{[r]}(\vec{\lambda})
{\rm cof}(\Gamma_{\mu}\lambda^{\mu}_{[r]}-M)W^0_{[r]}(\vec{\lambda})
\; ,
\label{explspin}
\eeq
where $N_{[r]}(\vec{\lambda})$ is a normalization factor
and $W^0_{[r]}(\vec{\lambda})$ any spinor only constrained by
$W^0_{[r]}(\vec{\lambda})\not\in{\rm Ker}[{\rm cof}
(\Gamma_{\mu}\lambda^{\mu}_{[r]}-M)]$.
The remaining spinors can be determined, e.g.,
with the symmetries discussed earlier.
If both the negative- and the positive-valued roots are degenerate
and an additional conserved quantity
commuting with the Hamiltonian is unknown,
no orthonormal spinor basis spanning the eigenspaces is preferred.
One can then replace ${\rm cof}(\cdot)$
with $\widehat{\rm cof}(\cdot)$ in Eq.\ (\ref{explspin})
and in the associated requirement on $W^0_{[r]}(\vec{\lambda})$.
It is now possible to proceed as in the nondegenerate case.

\section{Summary}
\label{sum}

This work has discussed
the theory of the Lorentz-violating Dirac equation in the SME.
The main results include  various symmetry properties of the solutions
and generalizations of conventional relations.
In particular,
Eq.\ (\ref{aexpansion}) permits the construction
of a hermitian Hamiltonian to arbitrary order
in the Lorentz-violating parameters.
Symmetries of the eigenenergies and the eigenspinors
arising from the discrete C, P, and T transformations
are given by Eqs.\ (\ref{csymmetry2}) and (\ref{PT3}),
and by Eqs.\ (\ref{cspinor2}) and (\ref{ptspinor2}), respectively.
The analog of the conventional spinor projectors
is provided by Eq.\ (\ref{projnond}) in the non-degenerate case
or Eq.\ (\ref{projd}) in the degenerate case.
These latter two equations involve
the matrix of cofactors
of the modified Dirac operator,
which is given explicitly by Eq.\ (\ref{cofdef2}).

\acknowledgements
This work was supported in part
by the Centro Multidisciplinar de Astrof\'{\i}sica (CENTRA)
and by the Funda\c{c}\~ao para a Ci\^encia e a Tecnologia (Portugal)
under the grant POCTI/FNU/49529/2002.


\begin{references}
\bibitem{note2}
Note, however,
that the labeling
should reflect in a consistent way
the functional dependence of the roots
on the 3-momentum, the mass, and Lorentz-violating parameters.
To illustrate this issue, consider a $b_0$ model
with dispersion-relation roots labeled
$\la^0_{+(\al)}(\vec{\la},m,b_0)=\sqrt{m^2+(|\vec{\la}|+(-1)^{\al}b_0)^2}$ and
$\la^0_{-(\al)}(\vec{\la},m,b_0)=-\sqrt{m^2+(|\vec{\la}|+(-1)^{\al}|b_0|)^2}$.
This labeling entails the undesirable feature
that the behavior of $(\al)$ under C
depends on the sign of $b_0$.
A different choice of labels,
such as $|b_0|\rightarrow b_0$ in $\la^0_{-(\al)}$,
would be more natural.
This is issue is particularly important
for the PT transformation.
\bibitem{note1}
In the ordinary Dirac case,
either helicity or spin projection
with respect to a given direction
in the rest frame of the particle
is usually employed to characterize
the degeneracy of the eigenenergies.
Note that under charge conjugation
both of these operators
change sign.\cite{iz}
For the present purpose
it is therefore irrelevant
whether helicity or spin-projection
labels are taken as the conventional ones.
\bibitem{note3}
Suppose this were not true.
In the conventional case,
the arguments of the spinors, $\vec{p}$ and $m$,
remain unchanged under the combined transformation of parity inversion
and time reversal.
This implies that there exist spinors $W$ that are proportional
to their PT image $-iC\gamma^5\gamma^0W$.
That this is impossible
can be verified by an explicit calculation
in Pauli-Dirac representation.
Thus the claim holds true in the conventional case for any labeling.
Assuming that in the present context the spinors depend smoothly
on the Lorentz-breaking parameters as they approach zero,
the result can be extended to the case of small Lorentz violation.
\end{references}
\end{document}